\documentclass{aa}  %[referee]
\usepackage{graphicx}
\usepackage[varg]{txfonts}
\usepackage{enumitem}
\usepackage{amsmath,amssymb}
\usepackage{booktabs}
\usepackage{xspace}
\usepackage{textcomp}
\usepackage{xcolor}	
\usepackage[euler]{textgreek}
\usepackage{dirtytalk}
\usepackage{caption}
\usepackage{subcaption}
\newcommand{\oiii}{[\ion{O}{III}]\,88\,\textmu m}
\newcommand{\cii}{[\ion{C}{II}]\,158\,\textmu m}

\begin{document}

   \title{To see or not to see a $z\sim 13$ galaxy, that is the question}

   \subtitle{Targeting the \cii\ emission line of HD1 with ALMA}
   
   \titlerunning{To see or not to see a $z\sim 13$ galaxy, that is the question}

   \author{Melanie Kaasinen\inst{1}, Joshiwa van Marrewijk\inst{1}, Gerg\"o Popping\inst{1}, Michele Ginolfi\inst{1,2}, Luca Di Mascolo\inst{3,4,5}, Tony Mroczkowski\inst{1}, Alice Concas\inst{1}, Claudia Di Cesare\inst{6,7,8}, Meghana Killi\inst{9,10}, Ivanna Langan\inst{1,11}}
    
    \authorrunning{M. Kaasinen, J. van Marrewijk, G. Popping, et al.}    

   \institute{European Southern Observatory, Karl-Schwarzschild-Str. 2, D-85748, Garching, Germany\\
              \email{melanie.kaasinen@eso.org}
          \and
          Dipartimento di Fisica e Astronomia, Universit\`{a} di Firenze, Via G. Sansone 1, 50019, Sesto Fiorentino (Firenze), Italy
          \and
           Department of Physics, University of Trieste, via Tiepolo 11, 34131 Trieste, Italy
          \and
          INAF - Osservatorio Astronomico di Trieste, via Tiepolo 11, 34131 Trieste, Italy
          \and
          IFPU - Institute for Fundamental Physics of the Universe, Via Beirut 2, 34014 Trieste, Italy
          \and
          Dipartimento di Fisica, Sapienza, Universit$\rm{\grave{a}}$ di Roma, Piazzale Aldo Moro 5, 00185, Roma, Italy
          \and
          INFN, Sezione di Roma I, Piazzale Aldo Moro 2, 00185 Roma, Italy
          \and
          INAF/Osservatorio Astronomico di Roma, Via di Frascati 33, 00078 Monte Porzio Catone, Italy
          \and
           Cosmic Dawn Center (DAWN), Jagtvej 128, 2200 Copenhagen N, Denmark
           \and
            Niels Bohr Institute, University of Copenhagen, Lyngbyvej 2, 2100 Copenhagen Ø, Denmark
          \and
           Univ Lyon, Univ Lyon1, Ens de Lyon, CNRS, Centre de Recherche Astrophysique de Lyon (CRAL) UMR5574, F-69230 Saint- Genis-Laval, France\\
         %     University of Alexandria, Department of Geography, ...\\
         %     \email{c.ptolemy@hipparch.uheaven.space}
         %     \thanks{The university of heaven temporarily does not
         %             accept e-mails}
             }
             
    % Orchids:
    % Luca: 0000-0003-3586-4485
    % Claudia: 0000-0003-1408-7373
    % Tony: 0000-0003-3816-5372

   \date{Accepted by A\&A 16/01/2023}

% \abstract{}{}{}{}{}
% 5 {} token are mandatory
 
  \abstract
   {Determining when the first galaxies formed remains an outstanding goal of modern observational astronomy. Theory and current stellar population models imply that the first galaxies formed at least at $z=14-15$. But to date, only one galaxy at $z>13$ (GS-z13-0) has been spectroscopically confirmed.}
   	%Yet, the highest redshift galaxy to have been securely confirmed remains GN-z11, at $z\sim 11$.
   {The galaxy `HD1' was recently proposed to be a $z\sim 13.27$ galaxy based on its potential Lyman break and tentative \oiii\ detection with ALMA. We hereby aim to test this scenario with new ALMA Band 4 observations of what would be the \cii\ emission if HD1 is at $z\sim 13.27$.}
   {We carefully analyse the new ALMA Band 4 observations and re-analyse the existing ALMA Band 6 data on the source to determine the proposed redshift.}
   {We find a tentative $4\sigma$ feature in the Band 4 data that is spatially offset by $1\farcs7$ and spectrally offset by $190$ km s$^{-1}$ from the previously reported $3.8\sigma$ `\oiii' feature. Through various statistical tests, we demonstrate that these tentative features are fully consistent with both being random noise features.} 
   {We conclude that we are more likely to be recovering noise features than both \oiii\ and \cii\ emission from a source at $z\sim 13.27$. Although we find no credible evidence of a $z\sim 13.27$ galaxy, we cannot entirely rule out this scenario.  Non-detections are also possible for a $z\sim 13$ source with a low interstellar gas-phase metallicity or ionisation parameter and/or high gas density. Moreover, the new continuum and line upper limits provide no strong evidence for or against a lower-redshift scenario. Determining where and exactly what type of galaxy HD1 is, will now likely require JWST/NIRSpec spectroscopy.}

                % stability of gas spheres
                % Techniques: image processing / interferometric / spectroscopic / image processing / imaging spectroscopy
                % Galaxies: individual: HD1
                % Galaxies: formation / evolution
                % Galaxies: high-redshift
                % Methods: data analysis / observational
                %}

   \maketitle
%
%________________________________________________________________

\section{Introduction}

 	Determining when the first galaxies formed remains an outstanding goal of modern observational astronomy, with important implications for theories of structure formation, baryonic physics, and cosmology. Simulations \citep[e.g.][]{2018MNRAS.480.4842C, 2020MNRAS.494.1071G,2022MNRAS.513.5621P} and observations \citep[e.g.][]{2018Natur.557..392H, 2020MNRAS.497.3440R} imply that the first galaxies were already present at $z\sim 14-15$, but confirming (or rejecting) their existence remains a challenge. At the time of first submitting this manuscript, the highest-redshift galaxy to have been confirmed was GN-z11 at $z\sim 11$ \citep{oesch2014,2016ApJ...819..129O,2021NatAs...5..256J}. At the time of publication, four $z>10$ galaxies have spectroscopically confirmed with the James Webb Space Telescope (JWST), the highest-redshift galaxy being GS-z13-0 at $z=13.2$ \citep{2022arXiv221204568C,2022arXiv221204480R}. 

 	To identify potential first galaxies, two main photometric methods have been employed. The first method has been to search for galaxies with a `break' (or `drop') in emission bluewards of Lyman alpha (Ly\textalpha), resulting from the absorption of UV photons by the neutral hydrogen in the early Universe \citep[e.g.][]{2016ApJ...819..129O}. For $z>10$ galaxies, this break lies in the near-infrared bands. Using the Hubble Space Telescope (HST), many such Lyman break candidates have been discovered \citep[e.g.][]{oesch2014,2016ApJ...817..120C,2019ApJ...883...99S} and, since the release of the first JWST/NIRCam data, many more, even higher redshift $z>10$ candidates have been identified in this way \citep[e.g.][]{Castellano2022,Harikane2022b,Naidu2022,Naidu2022b,2023MNRAS.519.1201A,2023MNRAS.518.6011D,2023ApJ...942L...9Y}. For the second main photometric pre-selection method, candidates have been identified based on their red Spitzer IRAC [3.6]-[4.5] \textmu m colour (referred to as the `IRAC excess'), proposed to arise from intense [\ion{O}{III}]\,\textlambda\textlambda 4959,5007 \AA\ and H\textbeta\ emission within the 4.5 \textmu m bandpass \citep{2016ApJ...823..143R}. Both of these photometric techniques are highly efficient in amassing high-redshift candidates, but additional spectroscopic observations are still required to confirm the proposed redshifts.

 	Spectroscopically confirming galaxies beyond $z\sim 7$ remains a challenge. The Ly\textalpha\ emission line, typically used at lower redshifts, is significantly attenuated for galaxies in the Epoch of Reionisation (EoR) by the high columns of neutral hydrogen surrounding the galaxies. Rest-frame UV lines are also challenging to observe, as shown in the follow-up of the rest-frame UV emission from GN-z11 with Keck \citep{2021NatAs...5..256J}. Prior to JWST, one of the most efficient methods of spectroscopically confirming high-redshift candidates has been via their \oiii\ and \cii\ line emission, which both remain well within reach of the Atacama Large sub-/Millimetre Array (ALMA) even for galaxies at $z>10$. Using ALMA, a few $z\sim 8-9$ galaxies with Spitzer and Hubble photometry have already been spectroscopically confirmed \citep{2017ApJ...837L..21L,2018Natur.557..392H,2019ApJ...874...27T,2020MNRAS.493.4294B}. For some of these confirmed $z=8-9$ galaxies, the best fit models to their rest-frame-optical to near-infrared (NIR) photometry imply that a significant stellar mass must already have been present at $z=14-15$ \citep[e.g.][]{2018Natur.557..392H, 2020MNRAS.497.3440R}. The existence of such large stellar masses at these redshifts is also corroborated by simulations of first galaxies \citep[e.g.][]{2018MNRAS.480.4842C,2022MNRAS.513.5621P}. While multiple ALMA DDT programmes have now been conducted with the aim of confirming/rejecting $z>12$ galaxy candidates identified using HST or JWST photometry, there are no conclusive results yet \citep{2022MNRAS.tmp.3515B,Popping2022,2022arXiv221008413Y}. Thus, the interstellar medium (ISM) of these youngest of galaxies is yet to be observed with ALMA. 

 	Early last year, \cite{2022ApJ...929....1H} presented tantalising evidence for a potential $z = 13.27$ galaxy. The authors first created a catalogue of potential $z\sim 13$ candidates, by searching the photometric data sets of the COSMOS and XDS fields, performing deblending corrections and searching for a Lyman break in the H band. After removing any foreground interlopers, \cite{2022ApJ...929....1H} identified two potential $z\sim 13$ Lyman Break Galaxies (LBGs). They followed up the first candidate, referred to as HD1, with ALMA Band 6 observations using four frequency tunings to cover the redshift range of $12.6 < z < 14.3$ for the targeted \oiii\ emission line. From these observations, a tentative $3.8\sigma$ line detection was reported at $\approx 237.8$\,GHz, which, if interpreted as \oiii\ emission, would correspond to a redshift of $z=13.27$. If confirmed, HD1 would be the highest redshift galaxy observed as of yet, existing at only $\sim 320$ Myr after the Big Bang. The existence of such a galaxy would imply little to no evolution in the UV luminosity function of galaxies from $z=8$ to $z=13$ (see \citealt{Bowler2020} and \citealt{2022ApJ...929....1H}), in contrast to predictions by galaxy formation models for the evolution of the UV luminosity function of galaxies at $z>10$ \citep[e.g.][]{2019MNRAS.483.2983Y, 2020MNRAS.499.5702B}.  As potentially one of the earliest known galaxies to form, HD1 thus presents an important benchmark for galaxy formation models and cosmological simulations.
  
	The tentative line detection and photometric redshift constraints for HD1 leave a large degree of ambiguity. As demonstrated in \cite{2022ApJ...929....1H}, the spectral energy distribution of HD1 can also be described by a $z\sim 4$ galaxy with a very low star formation rate. In that case, the $3.8\sigma$ line emission may instead correspond to the CO(9-8) or CO(10-9) line emission. Alternatively, the $3.8\sigma$ feature may simply be the result of random noise features. To distinguish between these possible scenarios, we conducted ALMA Band 4 observations, covering the frequency that would correspond to the \cii\ emission of the proposed $z=13.27$ galaxy. Not only would a \cii\ detection confirm the redshift of HD1, it would also provide useful constraints on its ISM properties (e.g. ionisation state and metallicity).

	In this paper, we describe these new ALMA observations and their implications. We present the ALMA Band 4 observations and data reduction as well as a consistent re-analysis of the Band 6 data obtained by \cite{2022ApJ...929....1H} in Sect.~\ref{sec:observations}. We present the data analysis in Sect.~\ref{sec:line_detection} and summarise the implications of our work in Sect.~\ref{sec:summary}. Throughout this paper, we use a flat $\Lambda$-CDM concordance model (H$_0 =$ 70.0 km s$^{-1}$ Mpc$^{-1}$, $\Omega_M =$ 0.30).

%__________________________________________________________________

\section{ALMA observations and data reduction}
	\label{sec:observations}

    \subsection{The new Band 4 observations} % (fold)
    	\label{sub:band_4_observations}
    
        To confirm or reject the proposed redshift of HD1, we obtained ALMA Band 4 observations through the ALMA Director's Discretionary Time (DDT) programme (2021.A.00008.S, PI: G. Popping), targeting the \cii\ emission that would correspond to the tentative \oiii\ detection. We used a setup with four spectral windows (SPWs) centred on 131.287 GHz, 133.164 GHz, 143.206 GHz, and 145.087 GHz. All SPWs except the one centred on 133.164 GHz were observed in Time Division Mode (TDM) with a bandwidth of 1.875 GHz (to obtain a continuum measurement), whereas the other one was set up using 240 channels with a 7.812~kHz spectral resolution and 1.875 GHz bandwidth (to target the \cii\ line). The observations were centred on RA 10$^h$01$^m$51.31$^s$, Dec +02$^d$32$^m$50.0$^s$ and were carried out on March 3 and 6, 2022. The total on-source time was 4.9 hours, with an average PWV of 4.746 mm.
        
        All data processing and calibration were performed with \texttt{CASA}, version 6.2 \citep{2007ASPC..376..127M}. We used the calibrated measurement sets generated by the observatory for the imaging of the observations. Imaging was performed with the \texttt{tclean} task, %the ALMA imaging pipeline\footnote{\url{https://almascience.eso.org/documents-and-tools/alma-science-pipeline-users-guide-casa-6-2.1}},  
        adopting natural weighting. We create both a continuum map (using the SPWs observed in TDM) and spectral cube (using the SPW centred at 133.16\,GHz). No continuum emission is detected. The continuum root-mean-square (rms) noise level is 5.2 \textmu Jy beam$^{-1}$ (yielding a $3\sigma$ upper limit of 15.6 \textmu Jy beam$^{-1}$). For the spectral cube at 133.16\,GHz, the rms is 127.3 \textmu Jy beam$^{-1}$ per 78 MHz. The continuum and spectral cube image have a beam of $2\farcs78 \times 2\farcs 16$ and $2\farcs91 \times 2\farcs27$, respectively. The final data cube used in the analysis reported here spans a frequency range of $132.22-134.08$ GHz, corresponding to a redshift range of $z = 13.17-13.37$. The \textsc{CLEAN}ed Band 4 continuum map showed several off-centre point sources. For the spectral cube, these were removed through \texttt{CASA-imcontsub}\footnote{\url{https://casa.nrao.edu/docs/taskref/imcontsub-task.html}} (because \texttt{CASA-uvcontsub}\footnote{\url{https://casa.nrao.edu/docs/taskref/uvcontsub-task.html}} is less optimised for removing off-centre co-detections). 

% subsection band_4_observations (end)
    \begin{figure*}
	\centering
    \includegraphics[width=0.98\linewidth]{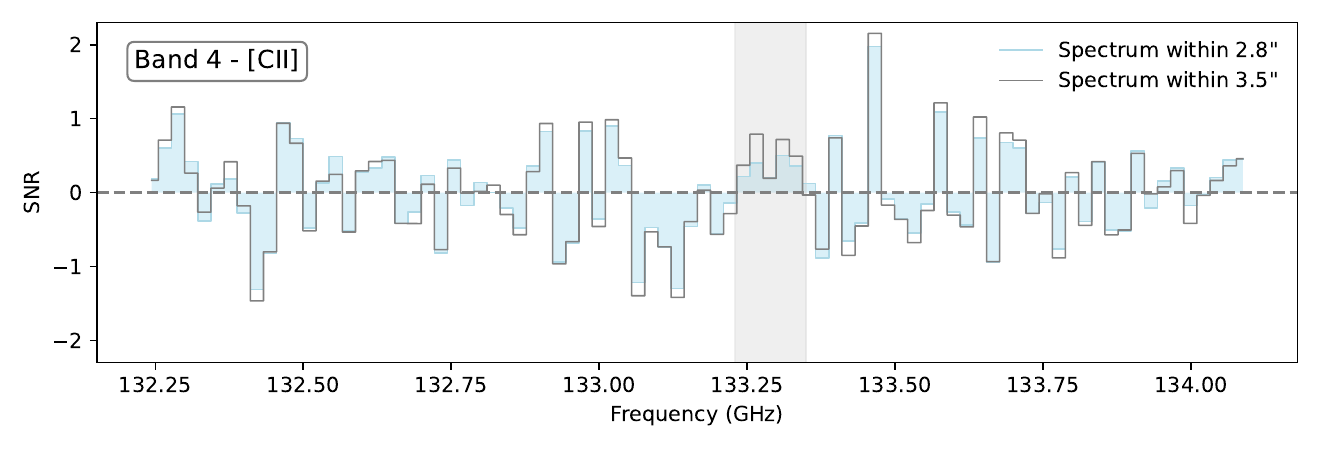}
    \includegraphics[width=0.98\linewidth]{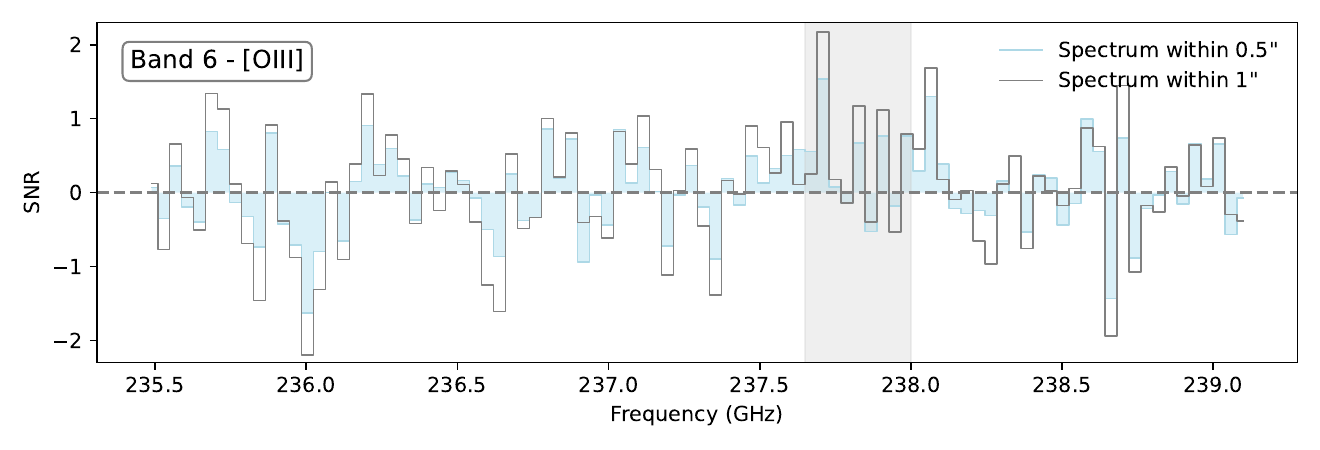}

	\caption{\label{fig:spectra} Integrated aperture spectra for HD1 around the expected frequency of the \cii\ (top panel) and \oiii\ (bottom panel) emission lines. To easily compare the spectra extracted within apertures of different size, spectra are presented in terms of signal-to-noise ratio rather than actual flux density. For the \cii\ emission we adopt apertures of $2\farcs8$ and $3\farcs5$,  corresponding to a similar size to the beam and 1.25 times the beam of the ALMA band 4 data, respectively. In the bottom panel we adopt an aperture of $0\farcs5$ and $1\farcs0$, corresponding to roughly the beam size and the aperture adopted in \citet{2022ApJ...929....1H} for the same data, respectively. The grey shaded area marks the tentative \cii\ feature (top panel), and the location of the \oiii\ line feature  presented in  \citet{2022ApJ...929....1H}.}
	\end{figure*}

    \subsection{Revisiting the ALMA Band 6 observations} % (fold)
    	\label{sub:band_6_observations}
    
        As stated in \citet{2022ApJ...929....1H}, the ALMA Band 6 DDT observations (2019.A.00015.S, PI: A. K. Inoue) are observed with a spectral scan setup that should target the \oiii~emission line for a redshift range of $12.6<z<14.3$. The observations span the frequency range 222 - 250 GHz. For the full observational details we refer to \citet{2022ApJ...929....1H}.  %\textcolor{red}{As for the Band 4 data, we time-averaged the calibrated data set in bins of 30s. We didn't do selfcal on Band 6}.
        
        We tested whether manual flagging could significantly alter the tentative \oiii~like feature visible in the image-plane \citep[Fig. 5 in][]{2022ApJ...929....1H}. Because antenna DA41 had slightly elevated noise levels over the entire time domain, we tested the impact of completely flagging data from this antenna and found no clear difference between the moment-0 maps generated using the initial, \texttt{CASA} version 6.3 pipeline reduced and calibrated data and those where we flagged antenna DA41. In this work, we therefore use the calibrated data reinstated using the ALMA pipeline generated scripts provided by the observatory. 
        
        We image the data using the \texttt{tclean} task, adopting natural weighting and creating cubes with channels of 50 km s$^{-1}$ width. The sensitivity per channel is 0.3 mJy beam$^{-1}$, where the beam FWHM is $0\farcs51$ $\times$ $0\farcs87$. We find no evidence of any continuum emission. For the continuum, the typical rms noise level is 8.0 \textmu Jy beam$^{-1}$, yielding a $3\sigma$ upper limit of 24.0 \textmu Jy beam$^{-1}$.

	\begin{figure*}
		\centering
		    \includegraphics[width=\textwidth, trim={0cm 1.3cm 0cm 1cm},clip]{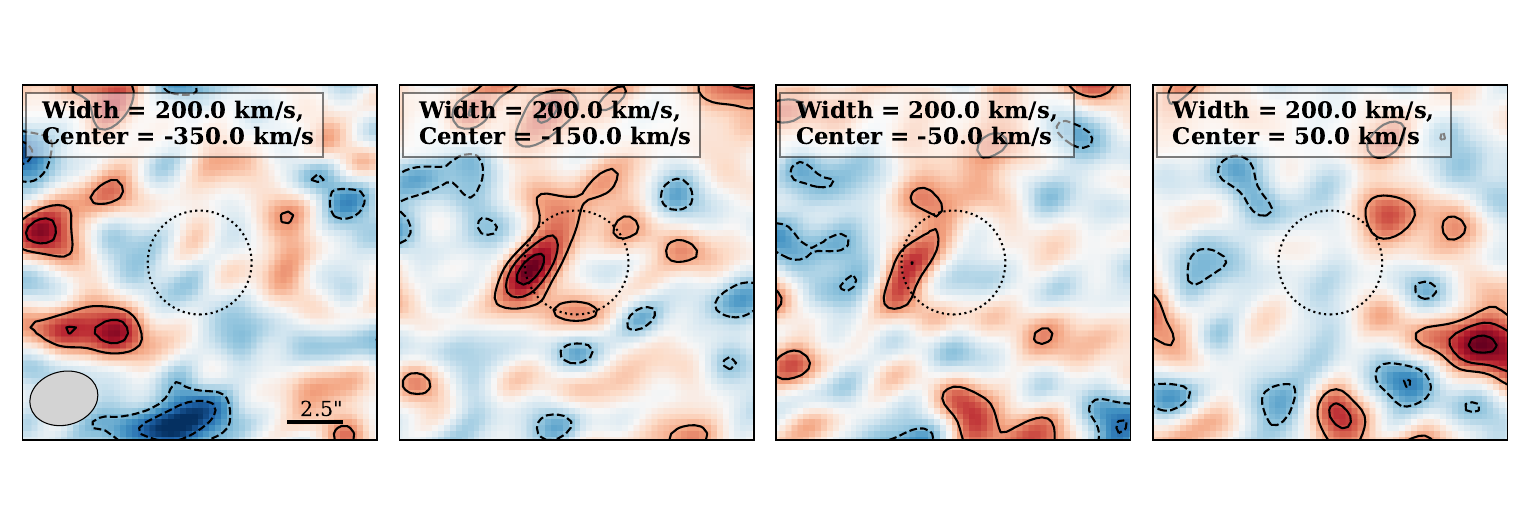}
		    \caption{\label{fig:mom0-Band4} ALMA Band 4 continuum-subtracted moment-0 maps of HD1. All maps are based on naturally weighted data. Each panel shows a moment-0 map collapsed over the same same integration width of 200 km s$^{-1}$, but centred at different frequencies, as indicated at the top of each panel. The central frequencies are indicated in km s$^{-1}$ with respect to the reference frequency $\nu = 133.2$\,GHz, the frequency of the expected [\ion{C}{II}] line. For all panels, the contours are drawn at -3.5,-2.5,-1.5, 1.5, 2.5, 3.5-$\sigma$ and have rms values of 8.32, 8.25, 8.44, and 8.88 mJy beam$^{-1}$ (from left to right). The synthesised beam FWHM is indicated by the ellipse in the lower left and the image scale is shown on the lower right in the left panel. The dashed circle has a $2\farcs2$ radius.} 
	\end{figure*}
	%\textcolor{red}{I actually didn't check if these panels overlap with atmospheric absorption lines. 

	\begin{figure*}
		\centering
		    \includegraphics[width=\textwidth, trim={0cm 1.5cm 0cm 1cm},clip]{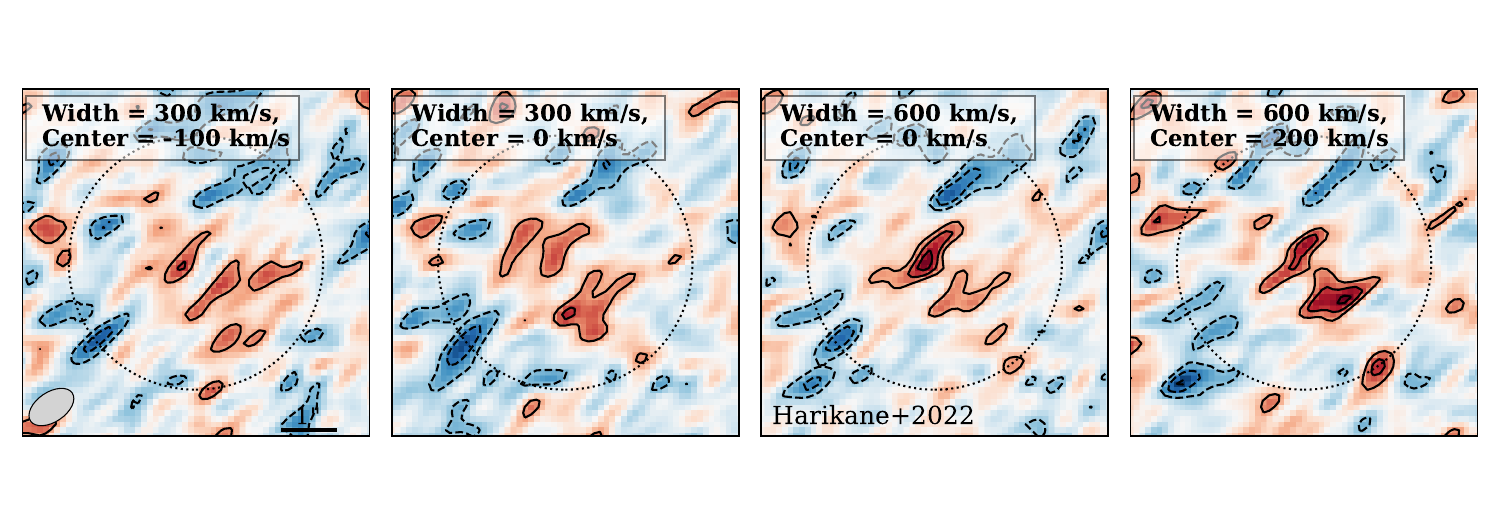}
		    \caption{\label{fig:mom0-Band6} ALMA Band 6 moment-0 maps of HD1. All maps are based on naturally weighted data. Each panel shows a moment-0 map created using a different integration width and central frequency. The integration width and central frequency are provided in km s$^{-1}$ with respect to the reference frequency, $\nu = 237.8$\,GHz, as reported in \citet{2022ApJ...929....1H}, at the top of each panel. For each panel, the contours are drawn at -3.5,-2.5,-1.5, 1.5, 2.5, 3.5-$\sigma$. From left to right, the moment-0 maps have rms values of 38.3, 37.4, 51.5, and 52.5 mJy beam$^{-1}$. The synthesised beam FWHM is indicated by the ellipse in the lower left and the scale of the image is shown on the lower right in the left panel. The dashed circle has a $2\farcs2$ radius.} 
	\end{figure*}
	
% subsection band_6_observations (end)

\section{Data analysis} % (fold)
\label{sec:line_detection}

% \begin{flushright}
% \textit{``Though this be madness, yet there is method in't.''} \\
% \emph{Hamlet}, William Shakespeare
% \end{flushright}

\subsection{Inspecting the \cii\ and \oiii\ images and spectra}
\label{sub:data products}

	We searched for the \cii\ (and \oiii) emission line from HD1, with the aim of securely confirming its spectroscopic redshift. Thus, we created aperture integrated spectra around the spatial location of HD1, focusing on the spectral range where the \cii\ and \oiii\ emission lines are expected, given the redshift proposed by \citet{2022ApJ...929....1H}. If HD1 is indeed at $z=13.27$, the \cii\ emission line should be located at a frequency of 133.18 GHz. In the top panel of Fig.~\ref{fig:spectra}, we present the integrated spectrum of HD1 around this frequency within apertures of $2\farcs 8$ and $3\farcs 5$ diameter (corresponding to approximately the beam size and 1.25 times the beam size). No emission line is visible around 133.18 GHz for either aperture, indicating that the \cii\ emission line is not detected at the expected frequency and location. However, there does appear to be a tentative feature centred at a frequency of $\sim 133.27$ GHz (indicated by the grey shading in Fig.~\ref{fig:spectra}), particularly for the larger aperture. We refer to this as the `tentative \cii\ feature' in the remainder of the text and describe it again later in this section.

	In the bottom panel of Fig.~\ref{fig:spectra}, we aim to reproduce the \oiii\ detection reported in \citet{2022ApJ...929....1H}, by presenting the integrated spectrum calculated within multiple apertures around the location of HD1. We adopted apertures of $0\farcs5$ and $1\farcs$ around HD1, corresponding to about twice the beam size of the data and an aperture identical to the one adopted in \citet{2022ApJ...929....1H}, respectively. We reproduced the feature at $\sim237.8$ GHz when adopting an aperture of 1 arcsec (i.e. approximately six consecutive channels with a positive flux, though most of them with S/N $< 1$). However, we found that this feature was less pronounced when we adopted the $0\farcs5$ aperture.

	To further investigate the tentative features visible in the spectra and test their robustness, we created moment-0 maps focusing on various spectral ranges around the location of the expected emission lines (varying the central frequency and width of the velocity-range over which channels are collapsed). Fig.~\ref{fig:mom0-Band4} shows four, naturally weighted, continuum-subtracted moment-0 maps integrated over a width of 200 km s$^{-1}$ around the expected \cii~redshifted frequency (133.18 GHz) and location of HD1. We find a tantalising $\sim 4\,\sigma$ feature, offset by $\sim$-150 km s$^{-1}$ from the expected frequency and spatially offset $1\farcs8$ from the location of HD1 reported in \citet[second panel]{2022ApJ...929....1H}. However, similar features are found throughout the data cube (see for example the right panel in Fig.~\ref{fig:mom0-Band4}). 

	We were able to reproduce the findings of \citet{2022ApJ...929....1H} when following the same procedures to create moment-0 maps (collapsed over 600 km s$^{-1}$), as shown in the third panel of Fig.~\ref{fig:mom0-Band6}. However, features of similar significance to the one found by \citet{2022ApJ...929....1H} are present when slightly shifting the central frequency ($\sim 200$ km s$^{-1}$) from the original reported central frequency (right panel of Fig.~\ref{fig:mom0-Band6}). Moreover, integrating over a narrower velocity width (the left two panels, consistent with other high-redshift studies, e.g. \citealt{2018Natur.557..392H}) yields no clear positive signal at the reported location of HD1. The multitude of tentative features found in the moment-0 maps of both the Band 4 and 6 data just by shifting the central frequency and/or width over which channels are collapsed, illustrates the difficulty in interpreting tentative detections. A more robust analysis of the noise properties of the data is therefore necessary to provide actual meaning to these features.

  \subsection{Noise properties of the ALMA data} % (fold)
    \label{sub:noise_properties}

	\begin{figure*}[t]
		\centering
		    \includegraphics[width=\textwidth, trim={0cm 0cm 0cm 0cm},clip]{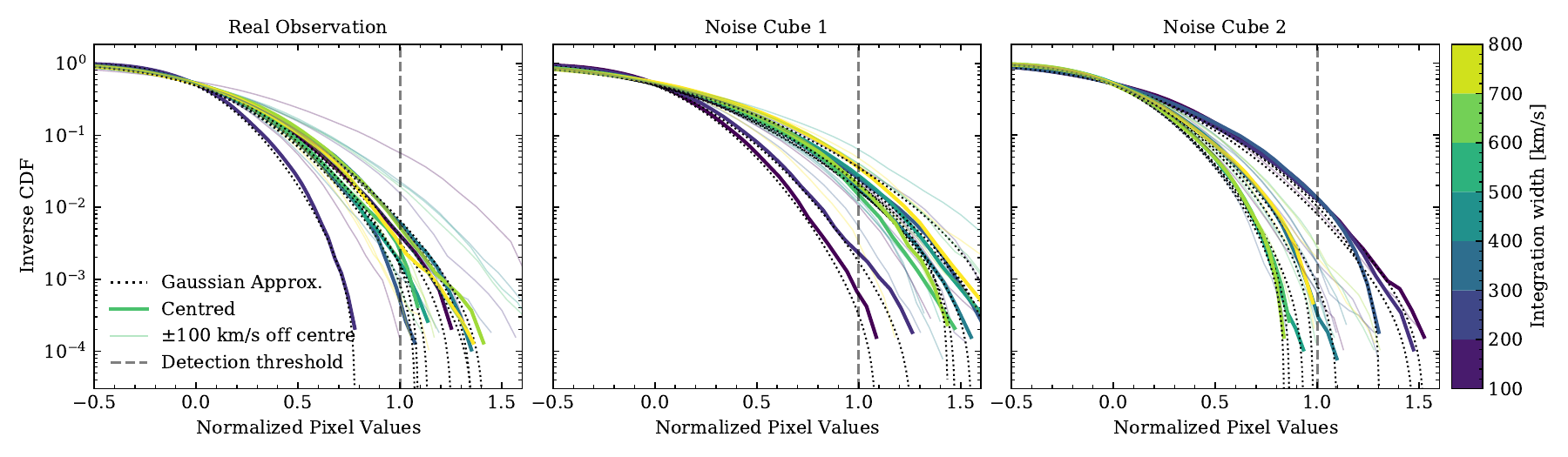}
		    \caption{\label{fig:hist-Band4} Inverse cumulative distribution of pixel values outside of a $5.2\arcsec$ aperture mask around the location of HD1, normalised by the brightest pixel value within that mask for different moment-0 maps of the ALMA Band 4 observations. Each moment-0 map is integrated with varying velocities widths (as indicated by the colourbar) and varying central frequencies that are off centred by $\pm$100~km s$^{-1}$ from the supposed \cii~line (which corresponding moment-0 maps are highlighted in bold). The left panel shows the real Band 4 observations, while the middle and right panel show the pixel-distribution of different pure noise observations (see Sect.~\ref{sub:noise_properties}). The thin dotted black lines are the assumed Gaussian distribution of the background fluctuations, estimated by computing the mean and standard deviation over the pixel values. Note that the brightest pixel value within the $5.2\arcsec$ aperture can be interpreted as the peak of the surface brightness of a source. Thus, distributions of pixels values that fall left of the vertical dashed line are indicative of a (tentative) detection within the aperture. For further clarification, see Sect.~\ref{sub:noise_properties}.}
	\end{figure*}
	
	\begin{figure*}[t]
		\centering
		    \includegraphics[width=\textwidth, trim={0cm 0cm 0cm 0cm},clip]{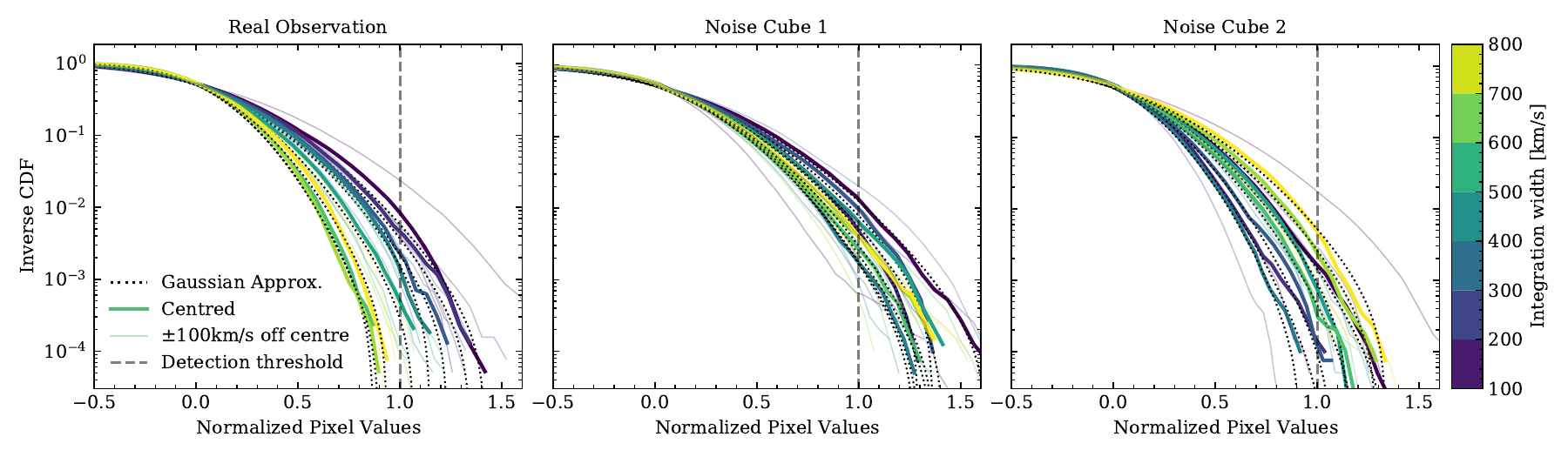}
		    \caption{\label{fig:hist-Band6} Same as Fig. 4 but for Band 6 and outside of a 2."2 aperture mask.}
  	\end{figure*}

  	In this section, we describe the tests we performed to determine whether the tentative line- and image-plane detections, in both the Band 4 and 6 data, could be attributed to structured noise or a non-Gaussian noise distribution. First, we compared the distribution of the pixel values (in flux density per beam) for different moment-0 maps, created with the Band 4 and 6 data (see Sect.~\ref{sub:data products}). The left panels of Fig.~\ref{fig:hist-Band4} and \ref{fig:hist-Band6} show the inverse cumulative distribution function (inverse CDF) of the signal-free pixel values for the Band 4 and Band 6 observations, respectively. We took the signal-free pixel values to be the pixel values that fall outside of a $2\farcs2$ radius for the Band 6 observations and a $5\farcs2$ radius for the Band 4 observations, centred on the phase reference position. Each line in Fig.~\ref{fig:hist-Band4} and \ref{fig:hist-Band6} indicates a unique moment-0 map that is integrated at a certain central frequency and over a certain integration width (with the latter indicated by the colourbar). In Fig.~\ref{fig:hist-Band4}, the `centred' lines, highlighted in bold, correspond to the velocity offset shown in the second panel in Fig.~\ref{fig:mom0-Band4} (-150 km s$^{-1}$). For Fig. \ref{fig:hist-Band6}, the bold lines correspond to the pixel distribution in the moment-0 maps that are centred at the reference frequency of the tentative \oiii~detection as reported in \cite{2022ApJ...929....1H}, and shown here in the third panel of Fig.~\ref{fig:mom0-Band6}. To guide the eye, the thin dotted black lines show the assumed Gaussian distribution of the background fluctuations, estimated by computing the mean and standard deviation over the pixel values per integration width from the highlighted moment-0 maps. 
    
    We found only minor deviations from the assumed Gaussian distribution (solid colour versus dotted black lines in Fig.~\ref{fig:mom0-Band6}), which could be caused by large-scale noise correlation (due to side lobes and beam smearing of the primary beam), imperfect continuum subtraction, gridding and weighting artefacts resulting from the discrete Fourier sampling of the visibilities, and/or minor atmospheric absorption. If the pixel distributions overshoot the Gaussian assumptions, these values would be associated with a higher significance than when compared to the actual underlying distribution. However, we did not detect any clear deviations that would indicate strong non-Gaussianity in the moment-0 maps.
    
    Finally, we normalised the distribution of pixel values by the brightest pixel value within the above described apertures. In the case of a real, unresolved source at the center of the map, this pixel value would be interpreted as the estimated peak source flux. Therefore, in Fig.~\ref{fig:hist-Band4} and \ref{fig:hist-Band6}, any CDF that ends at values lower than unity (leftwards of the dashed vertical line) implies a possible detection of a source within the central aperture, based on Gaussian statistics. Both left panels in Fig. \ref{fig:hist-Band4} and \ref{fig:hist-Band6} show that the tentative \cii~and \oiii~image-plane detections are only viable (left of the 1.0 mark) when integrated over a unique spectral range (200 and $\sim500-700$ km s$^{-1}$, respectively), which could give a physical insight into the velocity structure of HD1. We test how likely these features are to be real in the rest of this section and the following section. 
    
    % \subsubsection{Jack-knifed observations}% (fold)
    %     \label{sub:jack_knife}

    To effectively analyse the noise features we also created jack-knifed measurement sets. By randomly inverting the amplitudes of half of the visibilities and re-imaging each cube as for the original moment-0 maps, we created pure noise realisations, for which any possible signals were canceled out by the binning. This jack-knifing technique removed any true sources, while preserving the noise structure of the data. We repeated this jack-knifing procedure 10 times to generate 10 mock image cubes per ALMA Band, using different random seeds for the initial inversion in the $uv$-plane. The original and mock, jack-knifed data cubes have near equivalent noise properties and are all well-approximated by Gaussian distributions, as shown in the middle and right panel of Fig. \ref{fig:hist-Band4} and \ref{fig:hist-Band6}. 
        
    In Fig. \ref{fig:hist-Band4} and \ref{fig:hist-Band6}, we highlight two jack-knifed measurement sets per Band. The central panels show realisations wherein all the brightest pixel values fall outside of the central region ($2\farcs2$ for the Band 6 and $5\farcs2$ for the Band 4 data) for all moment-0 maps, which can be interpreted as a non-detection if the observations are treated as real. The right panels show two corresponding jack-knifed moment-0 maps per Band (with noise properties identical to the real observations) for which the brightest pixel value falls within the central region for specific integration widths when making the moment-0 maps. These noise peaks would thus be interpreted as a (tentative) detection when compared with the approximated Gaussian noise distribution. The middle and right panels that show the jack-knifed data cubes reveal very similar CDFs to the real data, with many pixels exceeding the source flux. The close match between the source-free jack-knifed cubes and real data for both Bands indicates that the tentative \cii~ and \oiii~lines are consistent with noise fluctuations.

      % subsubsection jack_knife (end)
  % subsection noise_properties (end)
  
  	\begin{figure*}[t]
		\centering
		    \includegraphics[width=0.49\textwidth, trim={0cm 0.2cm 0cm 0cm},clip]{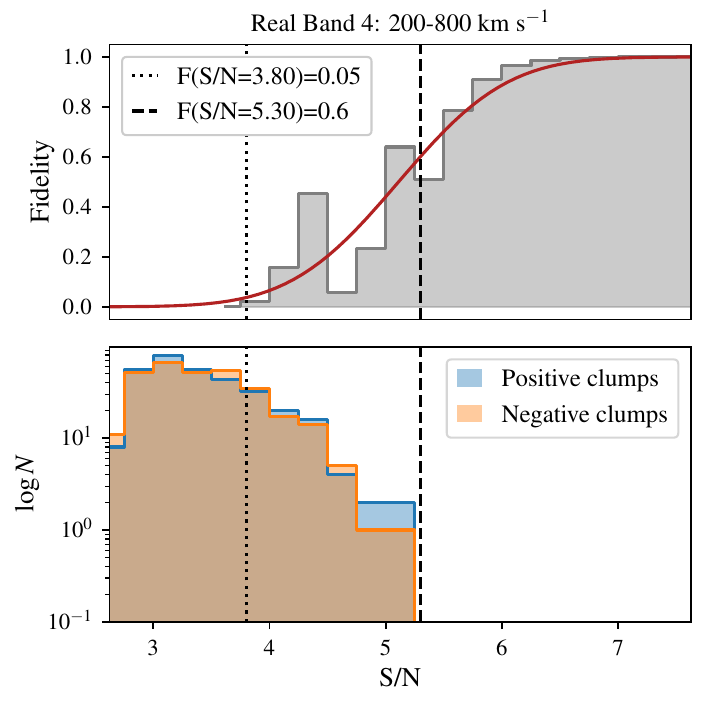}
		    ~
		    \includegraphics[width=0.49\textwidth, trim={0cm 0.2cm 0cm 0cm},clip]{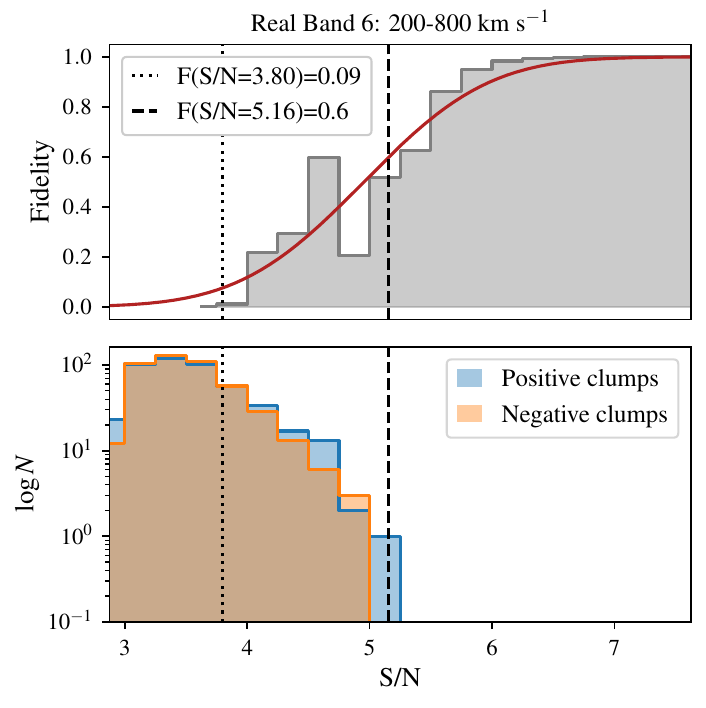}
		    \caption{\label{fig:fidelity} Fidelity (top) and number of peaks found (bottom) as a function of S/N for both the Band 4 data (left) and Band 6 data (right) using \texttt{Findclumps}. The dashed vertical line marks a fidelity of 0.6, beyond which it is more likely to select peaks that are positive than negative. The dotted vertical lines mark the S/N threshold of the tentative \oiii\ and \cii\ features. The red line in top panels is an error function fit to the fidelity histogram. This functional form has no underlying physical significance but was found empirically to provide a good fit \citep[see][]{2016ApJ...833...67W}.}
  	\end{figure*}

  \subsection{Likelihood of detecting two, correlated, noise peaks } % (fold)
    \label{sub:statistical_correlations_between_the_band_4_and_6_data}

	We tested the likelihood of finding two tentative detections in each data set, in the case that neither corresponds to a real source. In other words, we estimated how likely we are to obtain emission-line-like features at the same redshift and location on the sky if neither are actually line emission from the source. To this end, we applied two line-finding algorithms to both the real and jack-knifed data sets: (1) the python-based line-finding algorithm used by \citealt{2020A&A...643A...2B} (see their Sect. 6) and \citealt{2022NatCo..13.4574G} and (2) the algorithm \texttt{Findclumps}, described in \cite{2016ApJ...833...67W}, which is available as part of the python library, \texttt{Interferopy} \citep{interferopy}.\footnote{\url{https://interferopy.readthedocs.io/en/latest/}} For the real data, we used the dirty image cubes and replicated the noise properties in the jack-knifed cubes as described in Sec.~\ref{sub:noise_properties}. We ensured that all data cubes have the same spectral resolution of 50 km s$^{-1}$ but kept the spatial resolution of the original imaged cubes (i.e. cell sizes of $0\farcs1$ and $0\farcs25$, and beam sizes of $0\farcs51 \times 0\farcs87$ and $2\farcs91 \times 2\farcs27$, for the Band 6 and 4 data respectively). The frequency ranges of the data are the same as those shown in Fig. \ref{fig:spectra}; the Band 6 and 4 data sample similar redshift ranges for \oiii\ and \cii\ emission ($z=13.19-13.41$ and $z=13.17-13.37$ respectively).
   
   	The first line-finding algorithm that we adopted loops over all channels, excluding those at the edge of the cube (which have a higher rms), and, for each channel, creates moment maps of 200-800 km s$^{-1}$ velocity width. For each of these moment maps, the algorithm searches for pixels for which the absolute value exceeds that of the specified noise threshold (i.e. to find both positive and negative peaks), and for each of these central pixels, a spectrum is generated using a square aperture covering an area equivalent to that of the beam. If the S/N of the spectral feature, re-binned over the width of the moment-0 map, also exceeds the specified threshold, the feature is listed as a `peak'. \texttt{Findclumps} works slightly differently. Although it also collapses the cube over various user-specified `kernels', in our case integration widths of 200-800 km s$^{-1}$, it then runs \texttt{Sextractor} \citep{1996A&AS..117..393B} on the 2D image, to identify `clumps' in emission above the specified S/N. \texttt{Findclumps} never generates a spectrum to perform an additional S/N cut over, but uses only the 2D image.
   
   %Fixing the velocity resolution of the Band 4 data cube to 50 km s$^{-1}$ already reduced the significance of the potential \cii\ feature to $3.2\sigma$ in the moment map (and $3.1\sigma$ for the spectrum), implying that this line is far from being robustly detected. %We therefore use $3.1\sigma$ as the threshold to detect an emission feature in the following line-finding analysis.
   
   	Using both algorithms, we searched for lines with a velocity width between 200 and 800 km s$^{-1}$. We first performed this search for the entire cubes to test the significance of the tentative lines via the line fidelity, defined as,
	\begin{equation}
	    \mathrm{fidelity (S/N)} = 1 - \dfrac{N_\mathrm{neg} (\mathrm{S/N})}{N_\mathrm{pos} (\mathrm{S/N})}
	\end{equation}
	where $N_\mathrm{pos} (S/N)$ and $N_\mathrm{neg} (S/N)$ are the number of positive and negative line candidates with a given S/N, respectively \citep[see Sect. 3 of ][]{2016ApJ...833...67W} and the S/N is that of the peak in the velocity-integrated moment-0 image. In Fig. \ref{fig:fidelity}, we show the mean fidelity and number of detections per S/N bin (averaged over all integration widths) for the Band 4 and 6 data, as found with \texttt{Findclumps}. Using the first line-finding algorithm gave almost identical results. Following \cite{2016ApJ...833...67W}, we consider a fidelity threshold of 60\% as the minimum significance for a line detection (the vertical dashed line). The S/N corresponding to this fidelity threshold are 5.3 and 5.2, for the Band 4 and 6 data respectively. We repeated the same exercise for the jack-knifed data cubes, which are equivalent to pure noise cubes, any potential source having been nullified by the jack-knifing (see Sec.~\ref{sub:noise_properties}). We took the mean over all line widths and all 10 cubes. The mean S/N at a fidelity of 60\% for both the Band 4 and 6, jack-knifed cubes is $5.3\pm 0.2$. Conversely, the mean fidelity of the Band 4 and 6 jack-knifed cubes, at the S/N of the tentative features, are $0.06\pm 0.04$ and $0.09\pm 0.05$ respectively. The combined fidelity is then $= 0.5 \pm 0.4$ \%.  Thus, the significance of the tentative features, both on their own and in combination, is too low for them to be considered real.
   
   	Next, we tested the significance of the line detections given the expected source position. We searched for features within $\sim 2\farcs9$ of the expected source position, equivalent to $\leq 10$ kpc at $z=13.27$. This search region is motivated by the possibility of finding close companions as well as components within the exact same source. We based the S/N thresholds applied to both algorithms on the tentative \oiii\ and \cii\ lines. Since these features have a minimum moment-0 S/N threshold of $3.8\sigma$ (the \cii\ feature actually being at $4\sigma$) and spectral S/N threshold of $3.1\sigma$, we applied these cutoffs to the line-finding algorithms. We removed all duplicates (i.e. noise peaks with any spatial or spectral overlap), by keeping the component with the highest S/N. In this way, we generated a list of potential, positive and negative $\geq 3.8\sigma$ features (defined via the moment-0 S/N) of velocity width 200-800 km s$^{-1}$, for each cube. 
   
   	Applying the first line-finding algorithm to the real data cubes, we found one positive and one negative $\geq 3.8\sigma$ feature within $10$ kpc of the expected source position in the Band 4 data cube. We found three positive and eight negative $\geq 3.8\sigma$ features in the Band 6 data cube. Similarly, applying \texttt{Findclumps} to the Band 4 data cube, we found one positive and one negative $\geq 3.8\sigma$ feature within the $2\farcs9$ aperture. We also found five positive and nine negative $\geq 3.8\sigma$ features in the Band 6 data cube within this aperture. The slightly greater number of features found by \texttt{Findclumps} in the Band 6 data is due to the fact that \texttt{Findclumps} does not perform the additional S/N cut on the spectrum. All lists included the line candidates for the proposed $z=13.27$ source (by design). The higher number of $\geq 3.8\sigma$ features in the Band 6 data is mainly due to the difference in spatial resolution. The fidelity within this smaller search region is 0 for the Band 4 data but is negative for the Band 6 data.
   
    \begin{figure}[t]
		\centering
		    \includegraphics[width=0.49\textwidth, trim={0.5cm 0.5cm 0cm 0cm},clip]{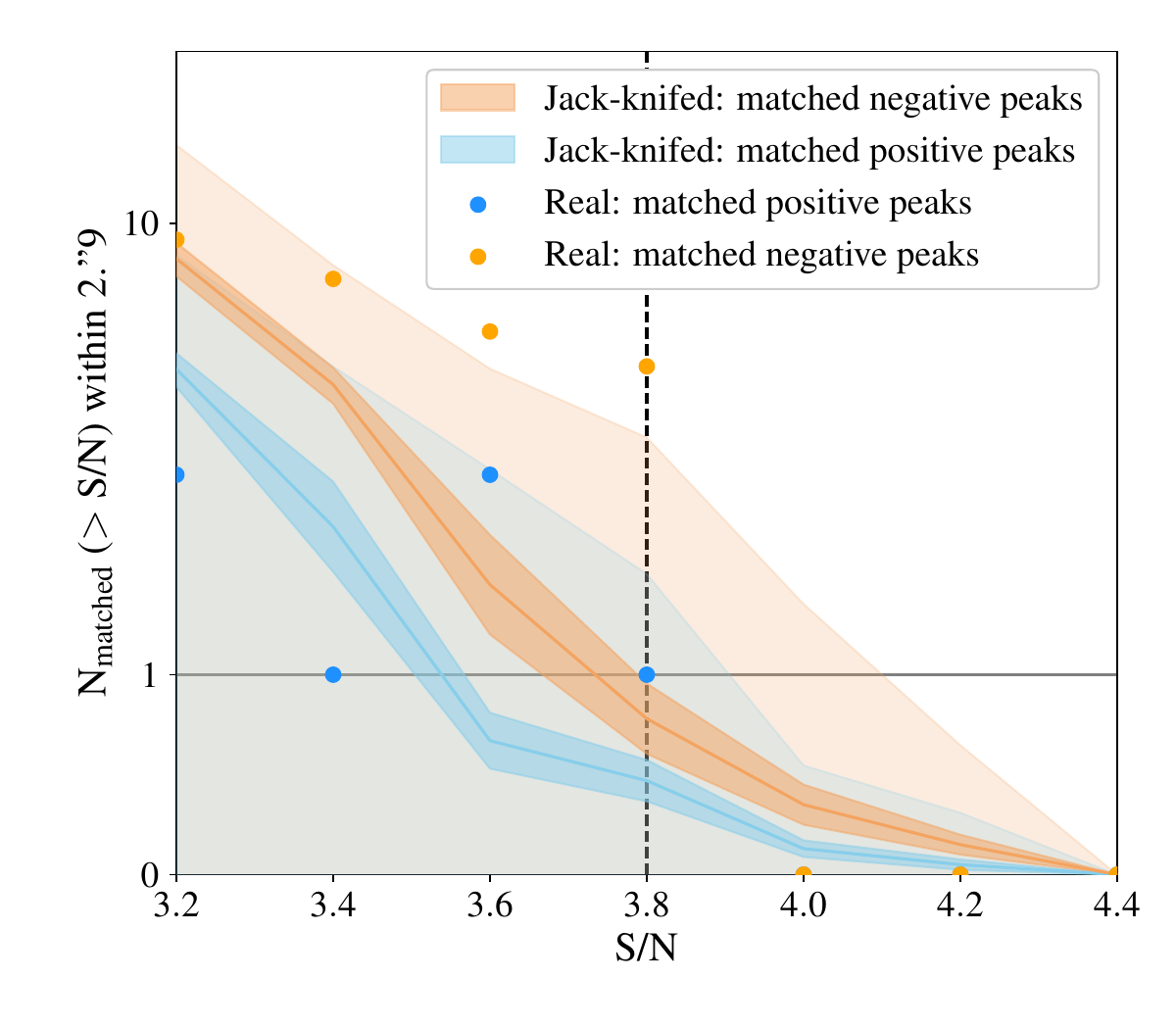}
		    \caption{\label{fig:matched_peaks} Number of matched peaks, within 1000 km s$^{-1}$, above a certain S/N and within $2\farcs9$ of the expected position of HD1. The filled blue and orange circles correspond to the number of positive and negative matched peaks, respectively, in the real data cubes. The solid line, dark- and light shaded regions correspond to the mean, standard deviation and standard error about the mean for the matched pairs of jack-knifed data cubes. We find one matched positive pair in the real data (the tentative \oiii\ and \cii\ features) at S/N$>3.8$. Given the mean and standard deviation on the matched peaks in the jack-knifed data, one set of matched peaks in the real data is perfectly consistent with being noise. }
  	\end{figure}
   
   	Cross-matching the real Band 4 and 6 data cubes, we recovered only one matched pair of positive $\geq 3.8\sigma$ features within 10 kpc and 1000 km s$^{-1}$ of each other, which correspond to the proposed \oiii\ line at $z=13.27$ (Band 6 cube) and tentative \cii\ line, offset by $\sim$ 6\,kpc and $\sim -190$\,km s$^{-1}$.\footnote{the slight difference in spectral offsets compared to the value in Sec.~\ref{sub:noise_properties} is due to the fact that here we are measuring the separation of the central line frequency from the line-finding algorithm.} However, we also found three matched pairs of negative $\geq 3.8\sigma$ features within 10 kpc and 1000 km s$^{-1}$ of each other, implying that the positive matched pair is unlikely to be real. The number of matched peaks are shown as the filled circles in Fig.~\ref{fig:matched_peaks}.
   
   	We also restricted the line-finding procedure to find peaks within $\sim 2\farcs9$ of the source position for the jack-knifed data cubes (of which there are 10 each for the Band 4 and 6 data). Of the 10 jack-knifed Band 4 cubes, five (four) have at least one positive $\geq 3.8\sigma$ feature within 10 kpc of the expected source position (the centre of the cube), whereas three (four) have at least one negative feature of the same significance. The two quoted values are for the two different line-finding algorithms. Of the 10 jack-knifed Band 6 cubes, \textit{all} have at least one positive $\geq 3.8\sigma$ feature within 10 kpc of the expected source position, with a mean of six (eight) such features per entire cube. Likewise, all jack-knifed Band 6 cubes have at least one negative $\geq 3.8\sigma$ feature, with a mean of seven (eight) such features per cube (with \texttt{Findclumps} finding the higher number of peaks).
   
   	We expanded this analysis, testing how likely we are to find two noise features (one each in the Band 4 and 6 data) that are offset by $\leq10$\,kpc and $\leq1000$\, km s$^{-1}$ for $3.2<\mathrm{S/N}<4.4$. The mean, standard error on the mean,\footnote{$\sigma/\sqrt{N}$, where $\sigma$ is the standard deviation and $N$ is the number of elements in the sample} and standard deviation (over all matched pairs) are shown in Fig. ~\ref{fig:matched_peaks}. Of the 100 pairs of matched Band 4 and 6 noise cubes, 30 pairs have at least one matching $\geq3.8\sigma$ noise feature, with a maximum of six matched noise peaks per pair of Band 4 and 6 noise cubes. The mean number of corresponding pairs of $\geq3.8\sigma$ emission features within 10 kpc and 1000 km s$^{-1}$ is 0.5, with a standard error on the mean of 0.1 and standard deviation of 1. Repeating the exercise for the negative peaks, we found that 25 of the matched cubes have at least one matched pair of negative peaks, with a mean of 0.8 such features, a standard error on the mean of 0.2 and standard deviation of 1.8. The large standard deviation for the matched noise cubes is the result of the small initial search area about the source centre. The mean number of matched peaks implies that with the data in hand there is, on average, a 50\% chance of detecting spatially and spectrally correlated $\geq 3.8\sigma$ noise peaks in both the Band 4 and 6 data. However, given the standard deviation, the likelihood is consistent with 100\%. We found that the number of matched noise peaks drops sharply with S/N. The number of jointly detected noise peaks drops to 0 at a S/N threshold of $\sim4.4$. Thus, at S/N$>4.4$ two matched positive features in the Band 4 and 6 data are certain to be real whereas at S/N$>3.8$ they are fully consistent with both being noise peaks.
   
  % subsection statistical_correlations_between_the_band_4_and_6_data (end)

% section line_detection (end)

% Two column figure (place early!)

\section{Exploring the potential redshift solutions}

  	\begin{figure*}[t]
		\centering
		    \includegraphics[width=\textwidth, trim={0cm 0.2cm 0cm 0cm},clip]{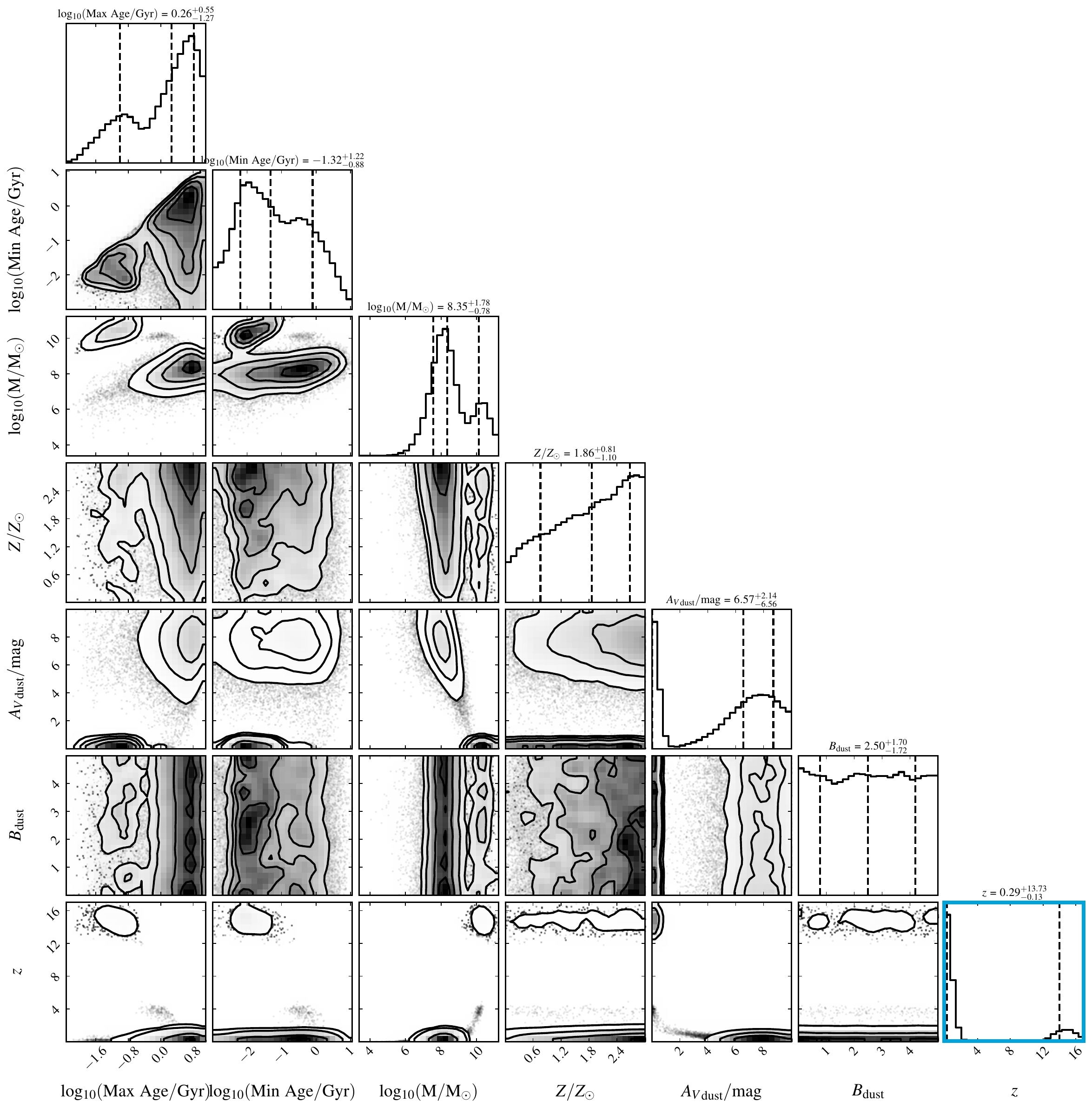}
		    \caption{\label{fig:modpipes_fit} Corner plot for the parameters fit to the photometry of HD1 using \texttt{Bagpipes}, assuming a constant SFH. From left to right (and top to bottom) posterior probability distributions are shown for the minimum and maximum age (i.e. the time since star formation switched off and on), the stellar mass ($M/M_\odot$), metallicity ($Z/Z_\odot$), optical extinction ($A_V$), 2175\,\AA\ bump strength ($B$) and redshift ($z$). With so few photometric detections, the 2175\,\AA\ bump strength and metallicity are unconstrained. As shown in the bottom right corner, the $z\sim 0.2-0.3$ solution dominates, with the $z\sim 14$ solution the next most probable. The low-z solution, which corresponds to a dwarf galaxy ($M_\star\sim 10^{8.3} \mathrm{M}_\odot$) with a very high optical extinction ($A_V\sim 7.5$) also dominates for other assumed SFHs (not shown here). However, the relative probabilities of these solutions are highly dependent on on the priors, as described in Sec.~4.3.}
  	\end{figure*}

  	\begin{figure*}[t!]
		\centering
		    \includegraphics[width=\textwidth, trim={0cm 0.5cm 0cm 0.2cm},clip]{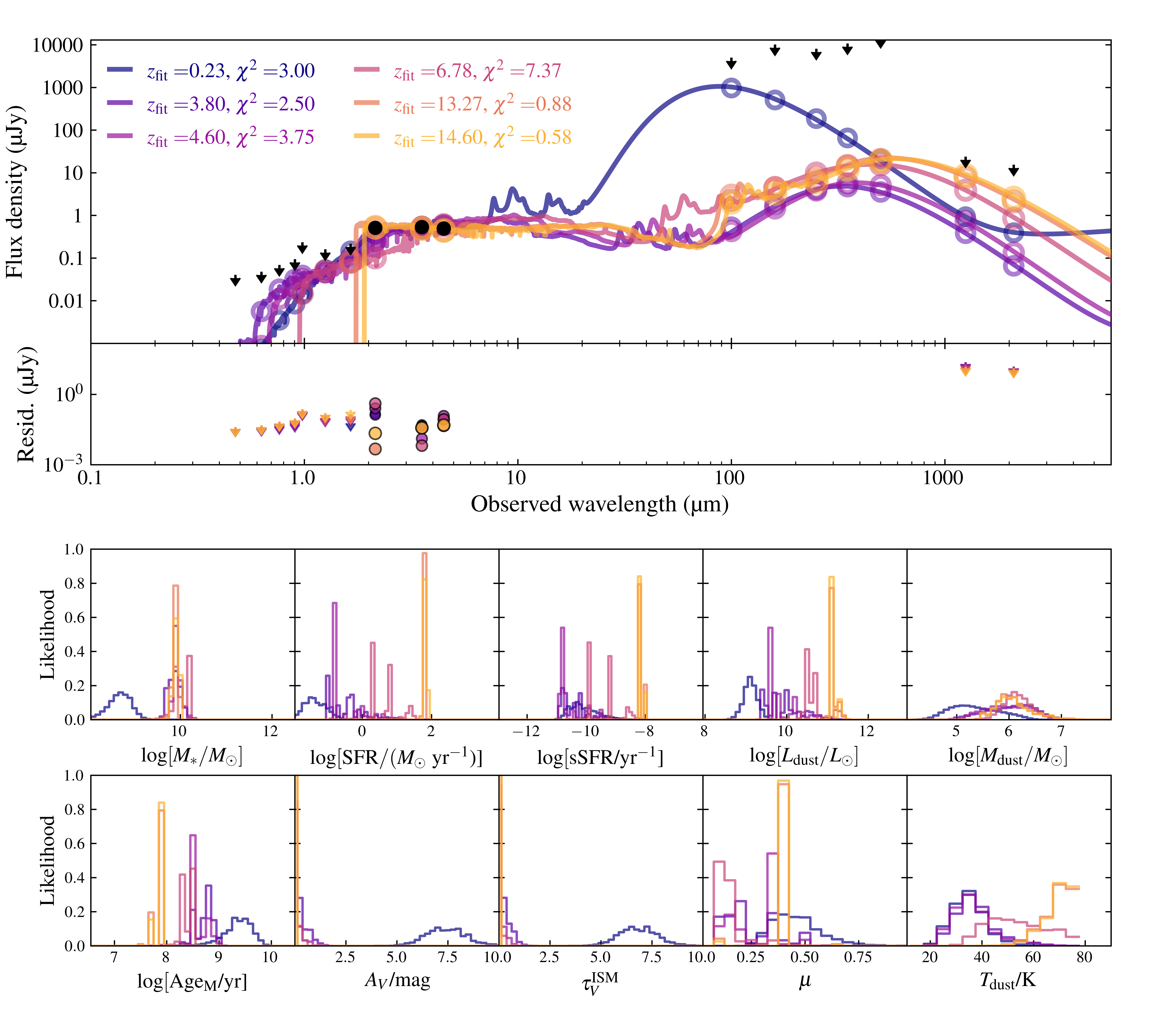}
		    \caption{Results of modelling the photometry for HD1 with \texttt{MAGPHYS}. Top panels: Best-fit SEDs for different potential redshift solutions, indicated by the legend at the top left, with the residuals shown below. The $z\sim 3.8$ and $z=4.6$ solutions correspond to a passive galaxy where the drop in flux is due to the Balmer break. The best intermediate-redshift fit to the limited available data is for $z=3.8$. In this scenario, the Band 6 data would cover CO(10-9), which we do not expect to be detected for such a passive galaxy (see text). The highest redshift solution within the range probed by the Band 6 data ($12.6<z<14.3$) provides the best fit overall. The single redshift solution between $z=4$ and $13$ for which the two $\gtrsim 3\sigma$ features could physically correspond to lines, $z=6.76$, is a much poorer fit to the photometry. Bottom panels: likelihood distributions for the different fit parameters, colour-coded according to the legend in the top panel.  \label{fig:sed_fits}}
	\end{figure*}

	Since we find no evidence of any line emission from a $z\sim 13.3$ galaxy, we revisit the likelihoods of the potential redshift solutions using the existing photometry and new upper limits. By modelling the spectral energy distribution (SED) of HD1 using the SED modelling tool \texttt{PANHIT}, \citealt{2022ApJ...929....1H} found two potential redshift solutions, $z\gtrsim 13$ and $z\sim 4$, corresponding to either a Lyman or Balmer break bluewards of 1.8 \textmu m. Since the new ALMA Band 4 data analysed here do not confirm the $z\sim 13.3$ scenario, we further explored other potential redshift solutions. We repeated the SED-fitting analysis to test the likelihood of different redshift solutions and checked whether these solutions are consistent with our upper limits on the submillimetre line emission (the continuum upper limits already being included in the SED modelling). To explore the likelihood of different redshift solutions, we used both the Bayesian Analysis of Galaxies for Physical Inference and parameter EStimation (\texttt{Bagpipes}; \citealt{2018MNRAS.480.4379C}) and the high-z version of the Multi-wavelength Analysis of Galaxy Physical Properties \citep[\texttt{MAGPHYS};][]{2015ApJ...806..110D,2019ApJ...882...61B} SED-fitting codes. Using these, we modelled the existing optical-NIR photometry (presented in \citealt{2022ApJ...929....1H}), the upper limits on the Herschel photometry,\footnote{available at \url{https://cosmos.astro.caltech.edu/page/herschel}} from the PACS Evolutionary Probe \citep[PEP,][]{2011A&A...532A..90L} and Herschel Multi-tiered Extragalactic Survey \citep[HERMES,][]{2012MNRAS.424.1614O} and the upper limits on the ALMA Band 6 and 4 continuum obtained here (Sec.~\ref{sec:observations}). 

\subsection{\texttt{BAGPIPES} fits}
    
	To model the SED of HD1 with \texttt{Bagpipes}, we followed a similar approach to \cite{2023MNRAS.518.6011D}. We applied the \cite{2018ApJ...859...11S} parameterisation to model the dust attenuation. For this, we centred the power-law deviation from the \cite{2000ApJ...533..682C} law to, $\delta=0$, applied a Gaussian prior with a standard deviation, $\delta=0.1$, and set a uniform prior on the 2175\,\AA\ bump strength, $B$, from 0 to 5. We set the logarithm of the ionisation parameter to $\log(U)=-3$ and allowed the stellar metallicity, $Z_\star$, to vary with a uniform prior from $0.01<Z_\star / Z_\odot<3$. We explored a variety of star formation histories (SFHs): constant, exponentially rising and double power law. For the constant SFH, we varied the time before observations at which stars began forming from 1 Myr to the age of the Universe. 

	To effectively exploit all the available spectral information, we incorporated the 14 upper limits in our modelling process in a way that is statistically meaningful in comparison to the three actual detections. We therefore extended the likelihood distribution function in \texttt{Bagpipes} to include an upper limit term that is consistent with the modified normal likelihood formalism for non-detections presented in \cite{2012PASP..124.1208S}. This implementation gave very similar results to setting all flux values to 0 in the standard version of \texttt{Bagpipes}. In this case, the $z>13$ solution has a far higher probability than the $z\sim 4$ solution, at least when we provide the wide and uniform priors described previously.  

	The likelihood distributions of the fit parameters (assuming a constant SFH) are shown in Fig. \ref{fig:modpipes_fit}. This results in two main redshift scenarios, $z\sim 0.3$ and $z\sim 14$, with the former appearing far more likely. To allow the nested sampling algorithm to even find the $z>0.3$ solutions, we needed to use a large number of line points (5000). In other words, when we left large uniform priors of $A_V= 0-10$ and $B=0-5$, we found that the most likely solution is neither of the two scenarios presented in \cite{2022ApJ...929....1H}, but instead a third scenario of a dwarf galaxy at $z\sim 0.2-0.4$, with stellar mass of $10^{8.3}\,\mathrm{M}_\odot$, $A_V\sim 6-10$ and a stellar population age of a few Gyr.

	The low-redshift scenario dominates for all SFHs, as long as we do not restrict the $A_V$. There is a difference of $\Delta z=0.05$ between the dominant, low-redshift solutions for a constant, exponential and double power law SFH. When we restricted the optical extinction further, for example to $A_V=0-5$, then the $z\sim 4$ solution started to become plausible. Further restrictions to $A_V$ ruled out the low-redshift solution entirely, increasing the likelihood of the passive $z\sim 4$ solution. When we forced the redshift range to $z=2-20$, but kept $A_V=0-10$ then the $z\gtrsim14$ solution dominated over that at $z\sim 4$. When restricting the redshift range in this way, the relative likelihood of the $z\sim 4$ and $z\sim 14$ solutions depended more strongly on the type of SFH and priors on the relevant ages. Given the range of redshift probability distributions that we found for different prior combinations, we are forced to conclude that with the available photometry there is no robust way to identify a single, best redshift fit. 

\subsection{\texttt{MAGPHYS} fits}

	We checked the \texttt{Bagpipes} results against \texttt{MAGPHYS high-z}, fitting for a range of potential redshift solutions, a subset of which are shown in Fig.~\ref{fig:sed_fits}. The redshift solutions that we chose to test are guided mainly by the results of the previous SED modelling performed by \cite{2022ApJ...929....1H} and here, with \texttt{Bagpipes}. We also fit for $z\sim 6.78$, a redshift solution that would correspond to a velocity offset of $1030$ km s$^{-1}$ between the CO(16-15) and CO(9-8) lines that would be covered by the Band 4 and 6 observations.

	We checked the best-fit SEDs for the different potential redshift solutions via the residuals and $\chi^2$ values of the best-fit \texttt{MAGPHYS} models (see legend at top left of Fig.~\ref{fig:sed_fits}). The $z\sim 0.3$, $z\sim 4$ and $z>13$ solutions, found with \texttt{Bagpipes} also provide a good fit when using \texttt{MAGPHYS}. The likelihood distributions of the stellar mass and $A_V$ for each redshift solution, are similar to those found using \texttt{Bagpipes}. In contrast to the model results using \texttt{Bagpipes}, solutions yielding a redshift of $z>13$ provide the best fit using \texttt{MAGPHYS}, in agreement with \cite{2022ApJ...929....1H}. This could be due to an array of factors, including the available model libraries or the different treatment of dust attenuation and emission, which are not worth exploring here. 

\subsection{Potential redshift solutions from the SED fitting}

	Given the range of redshifts that could be fit using \texttt{MAGPHYS} and \texttt{Bagpipes}, we conclude there are many possible redshift (and hence SFR, stellar mass etc) solutions that can be fit to the three photometric detections and 14 upper limits. Although the $z=6.78$ solution, for example, definitely provides a poorer fit, there is no good way to distinguish between $z\sim 0.3$, $z\sim 4$ and $z>13$ scenarios using the photometry alone. The SED fitting results are highly sensitive to any tiny colour offsets or calibration issues for the three photometric detections because the slope across these is too flat and they span a too narrow wavelength range for the best-fit solution to be trusted. For now, we conclude that there are three main possible sets of redshift solutions between which the current data cannot distinguish: (1) a dust-rich low-redshift ($z = 0.2-0.3$) dwarf galaxy, (2) a passive galaxy at $z = 3.6-4.6$, for which the photometric break corresponds to the Balmer break, or, (3) a $z>13$ galaxy, with a Lyman break bluewards of 2 \textmu m.

\subsection{Consistency between the SED fits and line upper limits}

	We explored whether the upper limits on the line luminosities covered by the Band 4 and 6 data are consistent with the three redshift solutions inferred previously from the photometry. For this, we used the mean rms per 50 km s$^{-1}$ channel, excluding the channels at the outskirts of the SPW, which is 0.319 mJy beam$^{-1}$ for the Band 6 data and 0.096 mJy beam$^{-1}$ for the Band 4 data. In each of the three cases, we assumed a line width of 200 km s$^{-1}$, applied a S/N threshold of 4, and assumed that the source is unresolved for both the Band 4 and 6 beam. The last assumption is motivated by the fact the maximum radius of the emission seen in the VISTA Ks image is $0\farcs7$, which is equivalent to the beam size of the Band 6 data and significantly smaller than the beam of the Band 4 data. This gives upper limits of 127.7 mJy km s$^{-1}$ and 38.4mJy km s$^{-1}$, for the Band 6 and 4 data respectively. 

\subsubsection{Low-redshift dwarf-galaxy scenario}

% Dwarf galaxies with low SFRs, old stellar pops?
% Little Things: https://ui.adsabs.harvard.edu/abs/2016AJ....151...14C/abstract
% SFRs ~ 10^{-2} Msun/yr
% Mstar ~ 0.6 x 10^8 Msun 
% [CII] dominant coolant, low dust temp
% littel extinction

% https://ui.adsabs.harvard.edu/abs/2017MNRAS.465.3741D/abstract
% 3 low-Z dwarf spheroidals
% M* ~ (3-10) x 10^8 Msun
% Mgas < 2 x 10^5 Msun
% GDR ~ 37-139 

	We first explored the $z\sim 0.2 - 0.3$ scenario, which corresponds to a dust-rich but barely star-forming (SFR\,$\sim 10^{-2}$ M$_\odot$ yr$^{-1}$) dwarf galaxy. In this case, a high visual extinction of $A_V\sim 7.5$ is required to explain the SED. For this scenario, the Band 6 observations would cover the frequency of HCN(3-2) for $z=0.06-0.20$ and CO(3-2) for $z=0.38-0.56$ whereas the Band 4 observations cover HCN(2-1) for $z=0.32-0.34$.

	Since HCN is typically far fainter than CO \citep[e.g.][]{2009A&A...496..677V}, we first tested whether we could have detected CO(3-2), for a dwarf galaxy at $z\sim 0.4$. The ALMA data provide an upper limit on the CO(3-2) line luminosity of $11.7 \times 10^{7}\, \mathrm{K \, km \, s^{-1} pc^2}$. The dust mass predicted by the \texttt{MAGPHYS} fit is $M_\mathrm{dust}\sim 10^{5.8}$ M$_\odot$. Supposing that the galaxy has a gas-to-dust mass ratio of 100, which is conservative for local dwarf galaxies \citep[e.g.][]{2017MNRAS.465.3741D}, the predicted gas mass is $6\times 10^7$ M$_\odot$. Assuming the galaxy follows the mass-metallicity relation of \cite{2020ApJ...891..181M}, the $\alpha_\mathrm{CO}$ vs metallicity relation of \cite{2012ApJ...747..124F} and assuming a conservative, Milky Way outer disk CO(3-2)-to-CO(1-0) line luminosity ratio of 0.3 \citep{1999ApJ...526..207F}, the predicted CO(3-2) luminosity is $\sim 3\times 10^6 \, \mathrm{K \, km \, s^{-1} pc^2}$. This estimate $37 \times$ lower than the upper limit on the line luminosity. Effects such as the variation in rms with frequency, minor changes in the dust-to-gas ratio or the galaxy actually having a higher metallicity cannot raise this value to more than the sensitivity. Thus, the lack of a line detection in the Band 4 and 6 data is perfectly consistent with the existence of a high-extinction, and low-SFR, dwarf galaxy at $z\sim 0.3$. Moreover, it may well be that the exact redshift is simply not covered by the observations. Although we cannot rule out this scenario, it has proven challenging to find an example of a low-redshift dwarf galaxy with such a high $A_V$ in the literature. %In saying that, there are also no confirmed $z>13$ galaxies yet. 

% de Looze+2017 Mdust~10**4 , GDR 37-139
% de Looze+2016 Mdust~5x10**3

\subsubsection{Passive $z\sim 4$ galaxy scenario}

	Next, we tested the case where the drop in flux bluewards of the observed-frame 1.8 \textmu m, corresponds to a Balmer break (i.e. drop in flux bluewards of the rest-frame 3646\,\AA). The potential redshift range consistent with the Balmer break solution is $z\sim 3.6 - 4.6$. For a subset of this range, $4.15<z<4.23$, the ALMA Band 4 observations would cover the frequency of the CO(6-5) line. The ALMA Band 6 observations would cover the frequency of CO(10-9) for $3.61<z<4.19$ and CO(11-10) for $4.07<z<4.71$. 

	We first considered the case of CO(6-5) emission from a z=$4.15$ galaxy, for which the upper limit on the line luminosity is $7.1\times 10^8\, \mathrm{K \, km \, s^{-1} pc^2}$. Converting the upper limit on the SFR from the best-fit solution found by \cite{Harikane2020}, SFR$< 0.1\,\mathrm{M}_\odot \mathrm{yr}^{-1}$, to an IR luminosity via the relationship of \cite{1998ARA&A..36..189K} (assuming a Chabrier IMF), gives $L_\mathrm{IR} = \mathrm{SFR} / 1.01 \times 10^{10} = 1 \times 10^9\, \mathrm{L}_\odot$. The best-fit value from \texttt{MAGPHYS} is $L_\mathrm{IR} = 3.6\times 10^9\, \mathrm{L}_\odot$. Given these $L_\mathrm{IR}$ vs $L_\mathrm{CO(6-5)}^\prime$ relation of \cite{2015ApJ...810L..14L}, these $L_\mathrm{IR}$ estimates translate to a predicted $L_\mathrm{CO(6-5)}^\prime = (0.4-1.2) \times 10^7\, \mathrm{K \, km \, s^{-1} pc^2}$, which is at least a factor of six below the CO(6-5) detection threshold. Thus, the lack of a line detection remains perfectly consistent with a passive galaxy at $z\sim 4.15$. 

	We repeated the same exercise for CO(10-9) at $z\sim 3.85$. Using the $L_\mathrm{IR}$ values predicted from the SED fits and $L_\mathrm{IR}$ vs line luminosity relations from \citealt{2015ApJ...810L..14L}), we found that the predicted value of $L_\mathrm{CO(10-9)}^\prime = (0.03-0.15) \times 10^7 \mathrm{K \, km \, s^{-1} pc^2}$ is far lower than the $4 \sigma$ line limit of $L_\mathrm{CO(10-9)}^\prime = 7.4 \times 10^8 \mathrm{K \, km \, s^{-1} pc^2}$. The CO(11-10) line is even fainter and thus even less likely to be detected. Since neither the lines nor continuum covered by the Band 4 and 6 observations for the passive $z=3.6-4.6$ galaxy solution should be detected at the sensitivity of our observations, we also cannot rule out (nor provide any stronger evidence for) this scenario.

\subsubsection{Star-forming $z>13$ galaxy scenario}

	Finally, we tested whether the lack of a significant line detection in the ALMA Band 4 or 6 data could rule out a galaxy at $z\sim 13.3$. In this case, the upper limits on the \oiii\ and \cii\ line luminosity are $5.2 \times 10^8 ~\mathrm{K \, km \, s^{-1} pc^2}$ and $4.9 \times 10^8~\mathrm{K \, km \, s^{-1} pc^2}$, respectively. In solar luminosity units, this corresponds to $L_\mathrm{[O \textsc{iii}]} < 6.4 \times 10^8~ \mathrm{L}_\odot$ and $L_\mathrm{[C \textsc{ii}]} < 1.1 \times 10^8~\mathrm{L}_\odot$. In their Fig. 11, \citet{Harikane2020} presented the predicted ratio of the \oiii\ and \cii\ luminosity with the SFR of galaxies based on \texttt{CLOUDY} \citep{Ferland2017} ionisation modelling. From the SED fitting performed here (and by \citealt{2022ApJ...929....1H}), the SFR for the $z\sim 13.3$ is $\sim 100\,\rm{M}\,\rm{yr}^{-1}$. Thus, the upper limits on these ratios are $\log{\mathrm{[\ion{O}{III}]/SFR}} < 6.7$ and  $\log{\mathrm{[\ion{C}{II}]/SFR}} < 6.0$. This is fully consistent with an ISM metallicity of $\leq 0.2\,\rm{Z}_\odot$ and a density of $\sim 10^3\,\rm{cm}^{-3}$. Moreover, \cite{2023MNRAS.520L..16K} recently conducted a more detailed study of simulated $z>10$ galaxies, showing that they mostly fall below the local metal-poor \oiii-SFR relation, mainly as a result of the low ionisation parameters ($U_\mathrm{ion} < 10^{-3}$). Even their [\ion{O}{iii}]-brightest simulated galaxy fell below the detection limit for HD1. Thus, the lack of any significant ALMA Band 4 or 6 detection corresponding to the \cii\ and \oiii\ emission lines, also does not rule out HD1 being a $z\sim13.3$ galaxy. It could simply imply a low ISM metallicity, low ionisation parameter and/or high gas density. Alternatively, if HD1 were at $z>14.3$ it would simply not have been covered by the Band 4 or 6 observations. 

% nearby source seen in Ks images is at z~0.81 (COSMOS 2020 catalogues)

\section{Summary} % (fold)
	\label{sec:summary}
	
	With this study, we aimed to confirm or reject the proposed redshift of $z=13.27$ for the galaxy, HD1. We recover the $3.8\sigma$ feature in the Band 6 data found by \cite{2022ApJ...929....1H} and find one $4\sigma$ feature in the Band 4 data that is spatially offset by $1\farcs7$ (6 kpc at $z\sim 13.27)$ and spectrally offset by 190 km s$^{-1}$. We quantify the significance of these features through various statistical tests, revisit the SED modelling and test whether the upper limits on the line luminosities are consistent with the best fit parameters from the SED modelling. We find that the tentative features in the Band 4 and 6 data are fully consistent with being noise peaks and thus find no credible evidence for a $z\sim 13.3$ galaxy. However, even with the new line and continuum upper limits, we cannot rule out this scenario nor lower-redshift, passive-galaxy scenarios. 

    By extensively testing the ALMA data, we determined the likelihood that the spatially and spectrally matched features found in the Band 6 and 4 cubes are consistent with noise peaks. To this end, we first performed a jack-knifing analysis, creating mock noise cubes with the same noise properties as the data, but with any real signal nullified. We find that the pixel distribution of these cubes closely matches that of the real data, with a significant number of pixels exceeding the value of any putative $z=13.27$ \oiii\ or \cii\ emission. That is, features at a statistical significance $\geq 3.8 \sigma$ (the significance of the proposed \oiii\ feature) appear often in the ALMA data, and should be treated with caution \citep[see also][]{2016A&A...589A..20V,2017A&A...604A.115V,2019A&A...627A.103V}. Moreover, by applying two line-finding algorithms to the real and jack-knifed data, we quantify the significance of the lines. We determine the line significance via the mean fidelity of the real and jack-knifed data, which is $6\pm 4 \,\%$ and $9\pm 5\,\%$ for the putative \oiii\ and \cii\ features, with a combined value of $0.5\pm0.4\,\%$. From this, and the significant number of positive and negative $\geq 3.8\sigma$ features detected within $2\farcs9$ (10 kpc at $z\sim 13.3$) of the source, we conclude that the putative features are likely noise peaks. Cross-correlating the positions of $\geq 3.8 \sigma$ noise peaks in the jack-knifed cubes, we find a mean of $0.5$ and standard deviation of 1 matched noise peaks (within $\leq 10$ kpc and $<1000$ km s$^{-1}$). Thus, even in combination, the $\geq3.8\sigma$ features recovered in the Band 4 and 6 data are fully consistent with being noise peaks. The likelihood that such matched features are consistent with noise only drops to 0 at S/N$=4.4$. 
    
    Revisiting the SED modelling, we found three possible redshift solutions, $z\sim 0.2 - 0.3$, $z\sim 4$ and $z>13$. These solutions all provide a good fit using \texttt{Bagpipes} and \texttt{MAGPHYS}. Applying these two different SED modelling tools, we find different likelihoods to \cite{2022ApJ...929....1H} and \cite{2022arXiv221103896F}, indicating that the results are highly sensitive to the input priors and underlying models. We caution against characterising the redshift probability using the existing photometry as the slope of the SED for the three photometric detections is too flat and they span too narrow of a wavelength range for the best-fit solution to be trusted. We test whether any of the three redshift solutions we found here can be ruled out by the upper limits on the lines covered by the ALMA observations. Even if it were within the redshift range for which the CO(3-2) would have been covered by the ALMA observations, we would not have detected the high-extinction and low-SFR dwarf galaxy that provided a good fit to the photometry. Likewise, the CO(6-5), CO(10-9) or CO(11-10) features that might have been covered for a passive $z\sim 4$ galaxy would not have been detected at the sensitivity of the observations. We also cannot rule out the $z>13$ scenario. For $z=13.17-13.37$, the redshift range covered by the ALMA Band 4 observations, it may simply be that the line luminosity is below the detection threshold due to a low ISM metallicity ($\lesssim 0.2~ \mathrm{Z}_\odot$), high gas density ($\gtrsim 10^3~\mathrm{cm}^{-3}$) and/or low ionisation parameter ($U_\mathrm{ion}\leq 10^{-3}$) \citep[as for the simulated $z>10$ galaxies in][]{2023MNRAS.520L..16K}. Moreover, given the wavelength coverage, we cannot rule out HD1 simply being at $z>13.37$ (i.e. the \cii\ line is not covered) or indeed $z>14.3$ (in which case neither the \oiii\ nor \cii\ line are covered by the existing ALMA data). Thus, we find that even with the addition of the ALMA Band 4 upper limits, the existing data cannot be used to constrain the redshift of HD1.
    
    Determining the redshift and properties of HD1 will require at least one of the following: (1) deep, sub-arcsecond resolution photometry at $\sim 100$ \textmu m (as the upper limits provided by Herschel provide too little constraining power), (2) $\gtrsim 10\times$ deeper observations at submillimetre wavelengths, or, (3) spectra covering the apparent break at $\sim 2$\,\textmu m and flux bluewards of the break. For (1), there is currently no suitable observational facility and (2) may require significantly more ALMA time (~ 500 h). Thus, option (3) using deep JWST/NIRSpec observations, will likely be the most feasible way to determine the nature of HD1. Indeed, recently four $z>10$ galaxies have been confirmed in this way \citep{2022arXiv221204568C}.
	
    Put in a broader context, we find that a secure \oiii\ or \cii\ detection with ALMA can be used to confirm a redshift, but for the typical sensitivities reached by ALMA in spectral scans, a non-detection cannot be used to rule out a redshift. On the other hand, once a high redshift has been confirmed, e.g. with JWST/NIRSpec spectroscopy, ALMA observations (including upper limits) may provide useful constraints on the metallicity and ionisation parameter of the ISM in the first galaxies. Looking to the future, we can expect significant advances to be provided by major new sub-/millimetre facilities, such as the Atacama Large Aperture Submillimetre Telescope (AtLAST; \citealt{Ramasawmy2022}). AtLAST will enable large, unbiased surveys of cosmological volumes in multiple bands, providing secure line identifications for large samples of high-$z$ galaxies without the need for photometric pre-selection \citep[e.g.][]{2023MNRAS.518.6142A}.  
    
    The main take-away message for the reader interested in confirming high-redshift galaxy candidates with ALMA, is to test the noise properties made in obtaining ALMA data products. All too often, we are willing to trust a $3-4 \sigma$ peak at the position where we expect to find it. But without testing how likely this is to be a noise fluctuation, we may well be sorely mistaken.

% section summary_of_implications (end)
\begin{acknowledgements}
	We thank the referee for a constructive report that helped improve the analysis and discussion of the likely redshift of HD1. We also thank Dr. Elisabete da Cunha and Dr. Leindert Boogaard for helpful discussions that greatly improved the quality of the analysis and interpretation of the data herein. Our thanks also to the creators of the publicly available codes used herein; thanks to Dr. Elisabete da Cunha for \texttt{MAGPHYS}, Dr. Adam Carnall for \texttt{Bagpipes} and Drs Mladen Novak, Leindert Boogaard and Romain Meyer for \texttt{Interferopy}. This paper makes use of the following ALMA data: ADS/JAO.ALMA\#2021.A.00008.S. ALMA is a partnership of ESO (representing its member states), NSF (USA) and NINS (Japan), together with NRC (Canada), MOST and ASIAA (Taiwan), and KASI (Republic of Korea), in cooperation with the Republic of Chile. The Joint ALMA Observatory is operated by ESO, AUI/NRAO and NAOJ. Calibrated measurement sets were provided by the European ALMA Regional Centre network \citep{Hatziminaoglou2015} through the calMS service \citep{Petry2020}. L.D.M. is supported by the ERC-StG `ClustersXCosmo' grant agreement 716762. AtLAST has received funding from the European Union’s Horizon 2020 research and innovation programme under grant agreement No. 951815.
\end{acknowledgements}   
\vspace{5mm}

\bibliographystyle{aa}
\bibliography{mybib}

\end{document}